\documentclass[letterpaper,prb,twocolumn,showpacs]{revtex4}
\usepackage[latin1]{inputenc}
\usepackage{ifpdf,ae}
\usepackage{amsmath}
\usepackage{amsfonts}
\usepackage{amssymb}
\usepackage{graphicx}
\usepackage{subfigure}

\begin{document}

\title{FMR and voltage induced transport in normal metal$-$ferromagnet$-$%
superconductor trilayers}
\author{Hans Joakim Skadsem and Arne Brataas}
\affiliation{Department of Physics, Norwegian University of Science and Technology,
NO-7491 Trondheim, Norway}
\author{Jan Martinek}
\affiliation{Institute of Molecular Physics, Polish Academy of Science, 60-179 Pozna\'n,
Poland}
\author{Yaroslav Tserkovnyak}
\affiliation{Department of Physics and Astronomy, University of California, Los Angeles,
California 90095, USA}

\begin{abstract}
We study the subgap spin and charge transport in normal 
metal-ferromagnet-superconductor trilayers induced by bias voltage  and/or
magnetization precession. Transport properties are discussed  in terms of
time-dependent scattering theory. We assume the  superconducting gap is
small on the energy scales set by the Fermi  energy and the ferromagnetic
exchange splitting, and compute the  non-equilibrium charge and spin current
response to first order in  precession frequency, in the presence of a
finite applied voltage.  We find that the voltage-induced instantaneous
charge current and  longitudinal spin current are unaffected by the
precessing  magnetization, while the pumped transverse spin current is 
determined by spin-dependent conductances and details of the  electron-hole
scattering matrix. A simplified expression for the  transverse spin current
is derived for structures where the  ferromagnet is longer than the
transverse spin coherence length.
\end{abstract}

\maketitle


\section{Introduction}

\label{sec:introduction}

Experimental and theoretical studies of spin polarized transport in hybrid
magnetic nanostructures is a frontier in mesoscopic physics. The most
prominent example of conceptual, technological, and commercial impact is the
giant magnetoresistance effect utilized in magnetic information storage
devices. In order to gain a deeper understanding of spin and charge
transport, and to enhance circuit functionality and efficiency, more complex
structures are fabricated and studied. In recent years, hybrid nanoscale
circuits containing normal conductors, ferromagnets, and superconductors
have been realized. These structures allow observation and understanding of
competing mechanisms of electron-electron interactions.

The simultaneous existence of ferromagnetism and superconductivity is rare.
In ferromagnets, the exchange interaction lifts the spin-degeneracy and
induces an itinerant spin polarization. In \emph{s}-wave superconductors, on
the other hand, electrons with anti-parallel spins form Cooper pairs. In
homogenous conventional ferromagnets (Fe, Ni, Co, and alloys thereof), the
large exchange interaction efficiently dephases electron-hole pairs, and
eliminates singlet superconducting correlations over distances larger than
the ferromagnetic coherence length. This would suggest a short-range
superconducting proximity effect in transition metal ferromagnets.\cite%
{kawaguchi92:_magnet_fe_nb,PhysRevB.55.15174} Such a simple picture cannot
explain recent measurements on Co and Ni ferromagnets coupled to Al
superconductors, however, where a substantial resistance drop was observed
at the onset of superconductivity.\cite%
{giroud98:_super_proxim_effec_in_mesos_ferrom_wire,petrashov99:_giant_mutual_proxim_effec_in}
The simple picture also fails to explain the long-range superconducting
proximity effect recently observed via the Josephson supercurrent through a
half-metallic ferromagnet.\cite%
{keizer06:_spin_tripl_super_throug_half,anwar10} Subsequent theoretical work
show that induced triplet superconducting correlations give rise to long
ranged proximity effect in transition metal ferromagnets.\cite%
{bergeret01:_long_range_proxim_effec_super_ferrom_struc,kadigrobov01:_quant}
Triplet superconducting correlations are insensitive to the pair-breaking
exchange interaction and exhibit a longer coherence length, similar to that
of superconducting correlations in normal metals. It is now established that
spin-flip processes in a ferromagnet can convert singlet into triplet pair
correlations. A spatially inhomogeneous magnetization texture \cite%
{bergeret05:_odd_tripl_super_and_relat} or magnons \cite%
{tkachov01:_subgap_trans_in_ferrom_super,takahashi07:_super_pumpin_in_josep_junct,houzet08:_ferrom_josep_junct_with_preces_magnet}
are examples of spin-flip sources that are able to induce long ranged
triplet correlations.

In this report, we focus our attention on the influence of magnons on the
transport properties in normal metal-ferromagnet-superconductor systems.
Even normal metal-ferromagnet systems without superconductors exhibit
intriguing physics, and especially the interaction between spin and charge
currents and the magnetic order parameter in such structures have attracted
tremendous interest. For instance, a non-collinear spin flow towards a
ferromagnet exerts a torque on the magnetization, a spin transfer torque,
that can excite the magnetization and even induce steady state, precessional
motion of the ferromagnetic order parameter.\cite%
{spintransfer2,spintransfer1} The inverse effect is also of significant
interest: A precessing ferromagnet in electrochemical equilibrium with its
environment, acts as a ``spin battery'' by emitting (or ``pumping'') pure
spin currents into neighboring materials.\cite{spinpumping} When emitted
spins are dissipated in adjacent materials, spin pumping enhances magnetic
dissipation in the precessing ferromagnet, and thus increases observed
linewidths in FMR experiments.\cite{mizukami01:_ferrom_nm_nm_nm_cu}

Some ideas from spin transfer physics in normal metal-ferromagnet structures
were recently used to study superconductor-ferromagnet systems. A FMR
experiment\cite{bell08:_spin_dynam_in_super_ferrom_proxim_system} and the
following theoretical analysis\cite{morten08:_proxim} have shown how spin
pumping can be used to visualize proximity effects and spin relaxation
processes inside the superconductor. In essence, in metallic contacts,
ferromagnetic correlations reduce the superconducting order parameter close
to the layer interface, enabling pumped sub-gap electrons to enter and
deposit spin in the superconductor. This is a prime example of how the
inverse proximity effect affects the FMR linewidth broadening when typical
spin-flip lengths are comparable to the superconducting coherence length.%
\cite{sillanpaeae01:_inver}

We direct our attention to a different aspect of the interplay between
magnetization and carrier dynamics in ferromagnet-superconductor structures.
In contrast to the works mentioned above, where the magnetization dynamics
have been the primary concern, we will consider how a precessing
magnetization and an applied voltage bias induce spin and charge currents in
a normal metal-ferromagnet-superconductor (N|F|S) trilayer. The computed
charge currents can be measured directly, whereas spin currents can possibly
be measured by its dissipative effect on the precessing ferromagnet, its
spin transfer torque effect on a second ferromagnet, or via spin-filtering
as a charge buildup on another ferromagnet.\cite{spinpumping} Related to our
work, sub-gap transport properties have recently been studied in a normal
metal-ferromagnetic superconductor structure.\cite%
{brataas04:_spin_and_charg_pumpin_by} In ferromagnetic superconductors,
magnetic and electron-hole correlations coexist which can result in novel
transport and dynamical magnetic phenomena. It was shown how superconducting
correlations, namely Andreev reflections at the layer interface, add
features to the results of spin and charge pumping in normal
metal-ferromagnet systems. In this report, we also consider how pumping in
the N|F|S trilayer is related to pumping in the normal metal-ferromagnetic
superconductor system as studied in Ref. %
\onlinecite{brataas04:_spin_and_charg_pumpin_by}.

Diffusive transport in hybrid superconductor-normal metal systems is usually
formulated within a quasiclassical description.\cite{rammersmith} Although
this description give qualitative insight into transport properties of
superconductor-ferromagnet systems,\cite%
{bergeret05:_odd_tripl_super_and_relat,houzet08:_ferrom_josep_junct_with_preces_magnet}
the formalism is limited to ferromagnets with exchange interactions much
smaller than the Fermi energy. Thus, a quasiclassical description cannot be
used to quantitatively study transport in transition metal ferromagnets Fe,
Ni and Co used in experiments. This is one of the reaons why we adopt the
scattering theory to transport.\cite{landauerbuttiker} Another reasone is
that scattering theory captures adiabatic slow time-dependent variations of
the magnetization direction well.

Scattering theory has proven most useful in the study of stationary charge
and spin currents in magnetoelectronic structures,\cite%
{brataas06:_non_collin_magnet} and the time-dependent generalization has
successfully been applied to describe parametric pumping of charge\cite%
{buettiker94:_curren,brouwer98:_scatt,buttiker06:_scatt_theor_of_dynam_elect_trans}
and spin currents.\cite{spinpumping} For the N|F|S structure under
consideration, we derive charge- and spin currents in the normal metal
conductor in response to a slowly precessing ferromagnetic exchange field
and applied bias voltage. We focus on sub-gap energies, and how Andreev
scattering contributes to the conductivites of the currents. In
electro-chemical equilibrium, we make contact with the results for pumping
in normal metal-ferromagnetic superconductor structures.\cite%
{brataas04:_spin_and_charg_pumpin_by} We proceed by detailing how time- and
energy gradients of the total scattering matrix contribute to
non-equilibrium pumped currents, and find that both charge and longitudinal
spin currents are unaffected by the precessing magnetization. Finally, we
consider non-equilibrium charge and spin currents for trilayers where the
ferromagnetic region is longer than the transverse spin coherence length.

This paper is organized in the following way: The N|F|S system is described
in Sec.~\ref{sec:model-description}. In Sec.~\ref{sec:time-depend-scatt}, we
use time-dependent scattering theory to derive general expressions for
charge and spin currents to first order in pumping frequency. The total
scattering matrix for the system is then invoked in Sec.~\ref%
{sec:evaluation-currents} to obtain non-equilibrium pumped currents. Our
conclusions are in Sec.~\ref{sec:conclusion}.

\section{Model description}

\label{sec:model-description}

\begin{figure}[ht]
\centering
\includegraphics[scale=0.32]{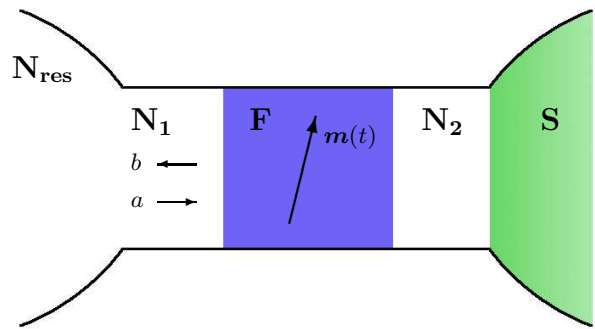}  \put(-140,80){\large{\textbf{F}}}
\put(-112,75){$\boldsymbol{m}(t)$}  \put(-30,80){\large{\textbf{S}}} 
\put(-75,80){\large{\textbf{N$_{\boldsymbol{2}}$}}}  \put(-185,80){\large{%
\textbf{N$_{\boldsymbol{1}}$}}}  \put(-230,100){\large{\textbf{N$_{\text{res}%
}$}}} \put(-185,65){$b$}  \put(-185,51){$a$}
\put(-160,67){\vector(-1,0){15}}  \put(-175,53){\vector(1,0){15}} 
\thicklines
\put(-125,45){\vector(1,4){10}}  
\caption{A ferromagnetic scattering region (F) is connected to a
superconductor (S) and a normal metal reservoir (N$_{\text{res}}$) via two
normal metal leads (N$_1$ and N$_2$). Amplitudes of outgoing (incoming)
carrier states are given by $b$ ($a$).}
\label{fig:1}
\end{figure}

The system is sketched in Fig.~\ref{fig:1}. It consists of a superconductor
(S) in series with a ferromagnet (F) and a normal metal lead (N$_1$). N$_1$
is ideally coupled to a normal metal reservoir (N$_{\text{res}}$). We assume
N$_{\text{res}}$ and S to be in local thermal equilibrium, and denote a
possible chemical potential difference between the normal and the
superconducting side as $\mu_{N} - \mu_{S} = eV$. Spin-orbit interactions
are disregarded, and the ferromagnetic order parameter is assumed to be
homogeneous and with a fixed magnitude $\Delta_{xc}$ inside F. Its direction
is along the time-dependent unit vector $\boldsymbol{m}(t) = (\sin \theta(t)
\cos \Omega t, \sin \theta(t) \sin \Omega t, \cos \theta(t))$. The
precessing magnetization serves as the pumping parameter in the system. 

We focus on sub-gap transport properties. Thus, possible scattering
processes include Andreev reflections at the F|S interface\cite{andreev64}
and spin-dependent normal scattering inside F. Following a standard
procedure,\cite{beenakker92:_quant_trans_in_semic_super_microj} the
scattering problem is greatly simplified by utilizing spatially separated
regions where scattering processes occur. This is achieved by inserting a
fictitious normal metal lead (N$_2$) between F and S. We assume that N$_2$
is longer than the Fermi wavelength, so that asymptotic, plane wave
solutions are applicable in this region. The total scattering matrix is a
concatenation of the scattering matrices for N$_1$|F|N$_2$ and for Andreev
reflections at the N$_2$|S interface. Transport between F and S is mediated
by the ballistic N$_2$ lead.

The singlet superconductor is described by the BCS Hamiltonian 
\begin{multline}  \label{eq:7}
\hat{\mathcal{H}} = \sum_{\sigma = \uparrow,\downarrow} \int\mathrm{d}%
\boldsymbol{r} \: \hat{\Psi}^{\dagger}_{\sigma} (\boldsymbol{r}) H_0 (%
\boldsymbol{r}) \hat{\Psi}_{\sigma}(\boldsymbol{r}) \\
+ \int \mathrm{d}\boldsymbol{r} \: \left\{ \Delta (\boldsymbol{r}) \hat{\Psi}%
^{\dagger}_{\uparrow}(\boldsymbol{r}) \hat{\Psi}^{\dagger}_{\downarrow}(%
\boldsymbol{r}) + \Delta^*(\boldsymbol{r}) \hat{\Psi}_{\downarrow}(%
\boldsymbol{r}) \hat{\Psi}_{\uparrow}(\boldsymbol{r}) \right\},
\end{multline}
where $H_0$ is the normal state, single-particle Hamiltonian and $\Delta(%
\boldsymbol{r})$ the superconducting gap. We model the gap by a step
function, $\Delta(\boldsymbol{r}) = \Delta\mathrm{e}^{\mathrm{i}\phi} \Theta
(x)$, where the phase $\phi$ is constant, and $x$ is the coordinate
perpendicular to the N$_2$|S interface. We take the Fermi energy $E_F$ to be
the largest energy scale, and focus on the low energy transport properties
in regimes when the superconducting gap is much less than the exchange
interaction in the ferromagnet $\Delta_{xc}$, $eV \le \Delta \ll
\Delta_{xc},E_F$. The Hamiltonian~(\ref{eq:7}) is diagonalized by the
following Bogoliubov transformation\cite{ketterson99:_super} 
\begin{equation}  \label{eq:8}
\hat{\Psi}_{\sigma}(\boldsymbol{r}) = \sum_n \left\{ \hat{\gamma}_n u_n (%
\boldsymbol{r},\sigma) + \hat{\gamma}^{\dagger}_n v^*_n (\boldsymbol{r}%
,\sigma) \right\},
\end{equation}
where $\hat{\gamma}_n^{(\dagger)}$ are quasiparticle annihilation (creation)
operators that satisfy the fermionic anti-commutation relation 
\begin{equation}  \label{eq:24}
\{ \hat{\gamma}_m, \hat{\gamma}^{\dagger}_{n} \} = \delta_{m,n}.
\end{equation}
The transformation~(\ref{eq:8}) results in a matrix equation for the
quasiparticle eigenfunctions $u_n$ and $v_n$: 
\begin{equation}  \label{eq:9}
\begin{pmatrix}
H_0(\boldsymbol{r}) & \mathrm{i}\Delta(\boldsymbol{r}) \sigma^y \\ 
- \mathrm{i}\Delta^*(\boldsymbol{r}) \sigma^y & - H_0^*(\boldsymbol{r})%
\end{pmatrix}
\begin{pmatrix}
u_n(\boldsymbol{r}) \\ 
v_n(\boldsymbol{r})%
\end{pmatrix}
= \varepsilon_n 
\begin{pmatrix}
u_n(\boldsymbol{r}) \\ 
v_n(\boldsymbol{r})%
\end{pmatrix}%
.
\end{equation}
The quasiparticle excitation energy $\varepsilon_n$ is measured with respect
to the chemical potential of the superconductor, which is set to zero. $%
\sigma^y$ is a Pauli matrix operating in spin space. The Bogoliubov-de
Gennes Hamiltonian~(\ref{eq:9}) is the starting point when we in Sec.~\ref%
{sec:scatt-matr-nfs} derive the appropriate reflection amplitudes for
quasiparticles impinging on the superconductor interface.

\section{Time-dependent scattering theory}

\label{sec:time-depend-scatt}

We now focus on the time-dependent scattering theory for the N|F|S structure
in Fig.~\ref{fig:1}, apply the general framework established in Refs.~%
\onlinecite{buettiker94:_curren,brouwer98:_scatt,vavilov01:_charg_pumpin_and_photov_effec,spinpumping,buttiker06:_scatt_theor_of_dynam_elect_trans}%
, and make use of the scattering theory for hybrid superconductor-normal
metal structures discussed in Refs.~%
\onlinecite{beenakker92:_quant_trans_in_semic_super_microj,blaauboer02:_charg_pumpin_in_mesos_system}%
. We find it most convenient to study a slowly precessing magnetization by a
scattering matrix expressed in the Wigner representation,\cite{rammersmith}
making the derivation of pumped currents similar to that carried out for
normal systems in Refs.~%
\onlinecite{wang02:_heat_curren_in_param_quant_pump,wang02:_optim_quant_pump_in_presen}%
.

In order to describe a scattering potential of arbitrary time-dependence, we
start by considering the two-time scattering matrix $\mathcal{S}(t,t^{\prime
})$, that relates annihilation operators between states outgoing and
incoming from the scattering region: 
\begin{equation}
\hat{b}_{\alpha }(t)=\sum_{\beta }\int \mathrm{d}t^{\prime } \mathcal{S}%
_{\alpha \beta }(t,t^{\prime })\hat{a}_{\beta }(t^{\prime }).  \label{eq:28}
\end{equation}%
As indicated in Fig.~\ref{fig:1}, $\hat{b}_{\alpha }:(\hat{a}_{\alpha })$
annihilates the outgoing (incoming) state $\alpha $. $\alpha $ labels
electron-hole Nambu space index, spin and transverse wave-guide number. We
assume that the reservoirs connected to the scattering region are in local
thermal equilibrium, and that incoming carriers from the normal metal
reservoir fulfill 
\begin{equation}
\langle \hat{a}_{\alpha }^{\dagger }(\varepsilon )\hat{a}_{\alpha ^{\prime
}}(\varepsilon ^{\prime })\rangle =\delta _{\alpha ,\alpha ^{\prime }}\delta
(\varepsilon -\varepsilon ^{\prime })f_{\alpha }(\varepsilon ),
\label{eq:29}
\end{equation}%
where the brackets indicate a quantum and statistical average, and 
\begin{multline}
f_{e(h)}(\varepsilon )=f_{0}(\varepsilon -\sigma ^{e(h)}eV)  \label{eq:5} \\
=\left[ 1+\mathrm{e}^{(\varepsilon -\sigma ^{e(h)}eV)/k_{B}T_{el}}\right]
^{-1},
\end{multline}%
where $\sigma ^{e(h)}=+(-)1$, and $f_{e(h)}(\varepsilon )$ is the
Fermi-Dirac distribution of incoming electrons (holes) at a charge bias $eV$
and electron temperature $T_{el}$. We will eventually consider electron
temperature to be lower than the superconducting gap. the We will now
proceed by computing charge and spin currents in the system.

\subsection{Matrix current}

\label{sec:matrix-current}

We seek the right-going charge and spin currents in normal metal lead 1, and
start by introducing the matrix current\cite%
{tserkovnyak01:_shot_noise_in_ferrom_normal_metal_system} 
\begin{equation}  \label{eq:20}
\hat{I}_{1,\alpha\beta}(t) = 2\pi e \tau^z_{\alpha\beta} \left( \hat{a}%
^{\dagger}_{\beta}(t) \hat{a}_{\alpha}(t) - \hat{b}^{\dagger}_{\beta}(t) 
\hat{b}_{\alpha}(t) \right),
\end{equation}
where $e$ is the electronic charge, and $\tau^z$ is a Pauli matrix in
electron-hole space: 
\begin{equation}  \label{eq:30}
\tau^z = 
\begin{pmatrix}
1 & 0 \\ 
0 & -1%
\end{pmatrix}%
.
\end{equation}
Charge and spin currents are obtained from the matrix current~(\ref{eq:20})
as follows: 
\begin{equation}  \label{eq:21}
I_{c} (t) = \sum_{\alpha} \langle \hat{I}_{1,\alpha\alpha}(t) \rangle,
\end{equation}
and 
\begin{equation}  \label{eq:22}
\boldsymbol{I}_s (t) = \frac{1}{2e} \sum_{\alpha,\beta} \boldsymbol{\rho}%
_{\alpha\beta} \langle \hat{I}_{1,\beta\alpha} (t) \rangle,
\end{equation}
respectively. Summations run over electron-hole, spin and mode space, and $%
\boldsymbol{\rho}$ is a matrix with diagonal structure in electron-hole
space: 
\begin{equation}  \label{eq:31}
\boldsymbol{\rho}_{\alpha\beta} \equiv 
\begin{pmatrix}
\boldsymbol{\sigma}_{\alpha\beta} & 0 \\ 
0 & \boldsymbol{\sigma}^*_{\alpha\beta}%
\end{pmatrix}%
,
\end{equation}
and with a vector of the Pauli matrices and their complex conjugates, as the
diagonal elements.

For a slowly oscillating scatterer, it is convenient to express the
scattering matrix in the Wigner representation\cite%
{rammersmith,wang02:_heat_curren_in_param_quant_pump,wang02:_optim_quant_pump_in_presen}
\begin{equation}
\mathcal{S}_{\alpha \beta }(t,t^{\prime })=\int_{0}^{\infty }\frac{\mathrm{d}%
\varepsilon }{2\pi }\mathrm{e}^{-\mathrm{i}\varepsilon (t-t^{\prime })}%
\mathcal{S}_{\alpha \beta }\left( \varepsilon ;\frac{t+t^{\prime }}{2}%
\right) .  \label{eq:32}
\end{equation}%
In this representation, the matrix current is: 
\begin{widetext}
  \begin{multline}
    \label{eq:81}
    \langle \hat{I}_{1,\alpha\beta}(t) \rangle = \frac{e}{2\pi}
    \tau^z_{\alpha\beta} \biggr\{ \delta_{\alpha,\beta}
    \int_0^{\infty} \mathrm{d}\varepsilon \: f_{\alpha} (\varepsilon)
    - \sum_{\gamma} \int_{-\infty}^{\infty} \mathrm{d}\tau \:
    \mathrm{d}T \int_0^{\infty} \frac{\mathrm{d}\varepsilon_1 \:
      \mathrm{d}\varepsilon_2}{2\pi} f_{\gamma}(\tau) \\ \times
    \mathrm{e}^{-\mathrm{i}\varepsilon_1 (T - \tau/2)}
    \mathrm{e}^{\mathrm{i}\varepsilon_2(T + \tau/2)}
    \mathcal{S}_{\alpha\gamma}\left (\varepsilon_2; t + \frac{T +
        \tau/2}{2} \right) \mathcal{S}^*_{\beta\gamma} \left(
      \varepsilon_1; t + \frac{T - \tau/2}{2} \right) \biggr\}.
  \end{multline}
\end{widetext}The current is expressed in terms of the center and relative
time coordinates $T=(t^{\prime }+t^{\prime \prime })/2$ and $\tau =t^{\prime
\prime }-t^{\prime }$, and the Fourier transform of the distribution
function 
\begin{equation}
f_{\gamma }(\tau )\equiv \int_{0}^{\infty }\frac{\mathrm{d}\varepsilon }{%
2\pi }\mathrm{e}^{-\mathrm{i}\varepsilon \tau }f_{\gamma }(\varepsilon ).
\label{eq:33}
\end{equation}

When the scattering matrix $\mathcal{S}(\varepsilon;t)$ is a concatenation
of multiple time-dependent scattering elements, the Wigner representation of 
$\mathcal{S}$ will also be an infinite sum of time and energy gradients.\cite%
{rammersmith} The magnetization dynamics is slow as compared to the time an
electron spends in the scattering region. In the \emph{adiabatic}
approximation, we assume the scattering matrix evolves on a much longer
timescale than the typical dwell times of particles inside the scattering
region. In this regime, we formally expand $\mathcal{S}$ as\cite%
{moskalets04:_adiab_quant_pump_in_presen} 
\begin{equation}  \label{eq:50}
\mathcal{S}(\varepsilon;t) = S_0(\varepsilon;t) + A(\varepsilon;t) + 
\mathcal{O}(\partial_t^2 S_0)
\end{equation}
where $S_0$ is the ``frozen'' or instantaneous scattering matrix, and the
matrix $A$ represents all first-order gradient corrections to $S_0$
resulting from the concatenation of time-dependent scattering elements that
describe the device. Unitarity of $\mathcal{S}$ to all orders in time- and
energy-gradients implies\cite{moskalets04:_adiab_quant_pump_in_presen} 
\begin{equation}  \label{eq:51}
S_0 A^{\dagger} + A S_0^{\dagger} = \frac{\mathrm{i}}{2} \left( \partial_t
S_0 \partial_{\varepsilon} S_0^{\dagger} - \partial_{\varepsilon} S_0
\partial_t S_0^{\dagger} \right) \equiv \frac{1}{2} P \left\{ S_0;
S_0^{\dagger} \right\} ,
\end{equation}
where a Poisson bracket $P\{.;.\}$ has been defined to ease the notation. In
the following, scattering matrix arguments $(\varepsilon;t)$ are omitted in
places where there is no risk of confusion.

To obtain a local (in time) expression for the matrix current~(\ref{eq:81}),
we Taylor expand $\mathcal{S}$ to first order in time derivatives, and
obtain the matrix current 
\begin{widetext}
  \begin{multline}
    \label{eq:82}
    \langle \hat{I}_{1,\alpha\beta}(t) \rangle = \frac{e}{2\pi}
    \tau^z_{\alpha\beta} \sum_{\gamma} \int_0^{\infty}
    \mathrm{d}\varepsilon \biggr\{ \left( f_{\alpha}(\varepsilon) -
      f_{\gamma}(\varepsilon) \right) \left( S_{0,\alpha\gamma}
      S_{0,\beta\gamma}^* + A_{\alpha\gamma} S^*_{0,\beta\gamma} +
      S_{0,\alpha\gamma} A^*_{\beta\gamma} - \frac{1}{2} P \left\{
        S_{0,\alpha\gamma}; S^*_{0,\beta\gamma} \right\} \right) \\
    + \frac{\mathrm{i}}{2} \left(- \partial_{\varepsilon} f_{\gamma}
      (\varepsilon)\right) \left( S_{0,\alpha\gamma} \partial_t
      S^*_{0,\beta\gamma} -
      \partial_t S_{0,\alpha\gamma} S^*_{0,\beta\gamma} \right)
    \biggr\} + \mathcal{O} \left( \partial_t^2 S_0 \right)
  \end{multline}
\end{widetext}where Eqs.~(\ref{eq:50})~and~\eqref{eq:51} have been used. The
matrix current in Eq.~(\ref{eq:82}) is \emph{exact} to first order in
frequency of the pumping parameter.

Finally, we observe that in the absence of a voltage bias, the gradient
corrections to the frozen scattering matrix, represented by $A$, vanish from
the matrix current. In electro-chemical equilibrium, when $V = 0$, $%
f_e(\varepsilon) = f_h(\varepsilon)$, the first line of Eq.~(\ref{eq:82})
vanishes, and the pumped current is determined by the frozen scattering
matrix. Naturally, the same is also true for the time-dependent theory based
on Floquet scattering matrices.\cite%
{buttiker06:_scatt_theor_of_dynam_elect_trans}

\subsection{Scattering matrix for a N|F|S structure}

\label{sec:scatt-matr-nfs}

In this section, the scattering matrix formalism derived for N|S structures%
\cite{beenakker92:_quant_trans_in_semic_super_microj} is applied to our
N|F|S trilayer. As described in Sec.~\ref{sec:model-description}, the
scattering description of a N|F|S structure is greatly simplified by
inserting a fictitious normal metal lead (N$_2$) between the two scattering
regions, thereby spatially separating spin-dependent scattering in F and
Andreev reflection at the N$_2$|S interface.\cite%
{beenakker92:_quant_trans_in_semic_super_microj} The scattering matrix $S_F$%
, describing the disordered ferromagnetic region, is block-diagonal in
electron-hole space. We write $S_F$ as 
\begin{equation}  \label{eq:34}
S_F(\varepsilon;t) = 
\begin{pmatrix}
s_F(\varepsilon;t) & 0 \\ 
0 & s_F(-\varepsilon;t)^*%
\end{pmatrix}%
,
\end{equation}
where the diagonal elements are 
\begin{equation}  \label{eq:35}
s_F = 
\begin{pmatrix}
r_{11} & t_{12} \\ 
t_{21} & r_{22}%
\end{pmatrix}%
.
\end{equation}
Here, $r_{ii}$ and $t_{ij}$ are matrices in spin-space that describe
reflection of an incoming electron in lead $i$, and transmission of an
electron from lead $j$ to lead $i$, respectively.

Electrons and holes with opposite spins are coupled by Andreev reflection at
the superconductor interface, where an incoming electron (hole) is reflected
as a hole (electron) with reversed spin direction. The reflection amplitudes
are derived by matching propagating wave functions in N$_2$ with evanescent
wave functions in the superconductor. The resulting scattering matrix reads%
\cite%
{beenakker92:_quant_trans_in_semic_super_microj,waintal02:_magnet_exchan_inter_induc_by_josep_curren}
\begin{equation}  \label{eq:36}
r^A = 
\begin{pmatrix}
0 & r^A_{eh} \\ 
r^A_{he} & 0%
\end{pmatrix}
= 
\begin{pmatrix}
0 & \mathrm{i}\alpha\sigma^y \mathrm{e}^{\mathrm{i}\phi} \\ 
-\mathrm{i}\alpha \sigma^y \mathrm{e}^{-\mathrm{i}\phi} & 0%
\end{pmatrix}%
,
\end{equation}
where $\alpha = \mathrm{exp} \left[ -\mathrm{i}\arccos (\varepsilon / 
\Delta) \right]$.

The total scattering matrix of the N|F|S structure is a concatenation of $S_F
$ and $r^A$, and in terms of the frozen scattering matrices, we obtain the
familiar results\cite%
{beenakker92:_quant_trans_in_semic_super_microj,blaauboer02:_charg_pumpin_in_mesos_system}
\begin{subequations}
\label{eq:6}
\begin{align}  \label{eq:16}
S_0^{ee} (\varepsilon&;t) = r_{11}(\varepsilon)  \notag \\
& + t_{12}(\varepsilon) r^A_{eh}(\varepsilon) r^*_{22}(-\varepsilon) M_e
(\varepsilon) r^A_{he}(\varepsilon) t_{21}(\varepsilon), \\
S_0^{hh} (\varepsilon&;t) = r_{11}^*(-\varepsilon)  \notag \\
& + t^*_{12}(-\varepsilon) r^A_{he}(\varepsilon) r_{22}(\varepsilon) M_h
(\varepsilon) r^A_{eh}(\varepsilon) t^*_{21}(-\varepsilon), \\
S_0^{eh} (\varepsilon&;t) = t_{12}(\varepsilon)
M_h(\varepsilon)r^A_{eh}(\varepsilon) t^*_{21}(-\varepsilon), \\
S_0^{he} (\varepsilon&;t) = t^*_{12}(-\varepsilon) M_{e}(\varepsilon)
r^A_{he} (\varepsilon) t_{21}(\varepsilon),
\end{align}
where time arguments are omitted on the right hand side of the equations for
sake of notation. Multiple reflections between S and F, mediated by
propagations through N$_2$, are described by 
\end{subequations}
\begin{align}  \label{eq:38}
M_e (\varepsilon) & = \left[1 - r^A_{he}(\varepsilon) r_{22}(\varepsilon)
r^A_{eh}(\varepsilon) r^*_{22}(-\varepsilon) \right]^{-1}, \\
M_h(\varepsilon) & = \left[1 - r^A_{eh}(\varepsilon) r^*_{22}(-\varepsilon)
r^A_{he}(\varepsilon) r_{22}(\varepsilon) \right]^{-1}.
\end{align}
From Eqs.~(\ref{eq:6}), and using $r^A_{eh}(-\varepsilon)^* =
r^A_{he}(\varepsilon)$, one obtains the following symmetry relations for the
total scattering matrix:  
\begin{subequations}
\begin{equation}  \label{eq:4}
\mathcal{S}^{ee}(\varepsilon;t) = \left[ \mathcal{S}^{hh}(-\varepsilon;t) %
\right]^*,
\end{equation}
and  
\begin{equation}  \label{eq:10}
\mathcal{S}^{eh}(\varepsilon;t) = \left[ \mathcal{S}^{he}(-\varepsilon;t) %
\right]^*.
\end{equation}

The frozen scattering matrices in Eqs.~(\ref{eq:6}) are all time-dependent
due to the slowly varying magnetization in the ferromagnet. Arguably the
easiest way to evaluate the matrix current, is to perform a spinor rotation
that aligns the spin quantization axis with the instantaneous magnetization
direction.\cite{spinpumping,brataas04:_spin_and_charg_pumpin_by} The total
scattering matrix 
\end{subequations}
\begin{equation}  \label{eq:40}
S_0 (\varepsilon;t) = 
\begin{pmatrix}
S_0^{ee} & S_0^{eh} \\ 
S_0^{he} & S_0^{hh}%
\end{pmatrix}%
\end{equation}
can be related to the total scattering matrix $\underline{S}$ in the
rotating frame by the spinor rotations 
\begin{equation}  \label{eq:25}
S_0 (\varepsilon;t) = {W}^{\dagger}(t) \underline{S} (\varepsilon) W(t),
\end{equation}
where $W(t) = V(t) U(t)$, with 
\begin{equation}  \label{eq:26}
U(t) = 
\begin{pmatrix}
\mathcal{U}(t) & 0 \\ 
0 & \mathcal{U}^{\dagger}(t)%
\end{pmatrix}
= 
\begin{pmatrix}
\mathrm{exp} \left[\frac{\mathrm{i} \Omega t}{2} \sigma^z\right] & 0 \\%
[0.3cm] 
0 & \mathrm{exp}\left[- \frac{\mathrm{i} \Omega t}{2} \sigma^z \right]%
\end{pmatrix}%
,
\end{equation}
and 
\begin{equation}  \label{eq:27}
V(t) = 
\begin{pmatrix}
\mathcal{V}(t) & 0 \\ 
0 & \mathcal{V}(t)%
\end{pmatrix}
= 
\begin{pmatrix}
\mathrm{exp}\left[\frac{\mathrm{i}\theta(t)}{2} \sigma^y\right] & 0 \\%
[0.3cm] 
0 & \mathrm{exp} \left[\frac{\mathrm{i}\theta(t)}{2} \sigma^y \right]%
\end{pmatrix}%
.
\end{equation}
In the rotating frame, $\underline{S}_0^{ee}$ and $\underline{S}_0^{hh}$ are
both diagonal in spin space, while $\underline{S}_0^{eh}$ and $\underline{S}%
_0^{he}$, which mix spin $\sigma$ electrons with spin $-\sigma$ holes, only
have off-diagonal elements.

Now that the matrix current and relevant scattering matrices are derived, we
proceed to study pumped charge and spin currents for a voltage biased
trilayer structure.

\section{Pumped currents out of equilibrium}

\label{sec:evaluation-currents}

A complication that arises when the system is driven out of equilibrium, is
that time- and energy gradients of the frozen scattering matrix must be
evaluated. Before presenting the detailed expressions for charge and spin
currents in the normal metal lead, we derive the required gradient
corrections. Due to electron-hole symmetry (\ref{symmetry}), it is
sufficient to consider only $A^{he}$ in the gradient correction.

\subsection{Gradient correction matrix}

\label{sec:grad-corr-matr-1}

In the following, we determine $A^{he}$ by a formal gradient expansion of
the corresponding scattering matrix $\mathcal{S}^{he}$, whose full time and
energy dependence of $\mathcal{S}^{he}$ is given by (see Eq.~(\ref{eq:16})): 
\begin{equation}  \label{eq:42}
\mathcal{S}^{he}(\varepsilon;t) = \left( t^*_{12} \circ M_e \circ r^{A}_{he}
\circ t_{21} \right)(\varepsilon;t).
\end{equation}
Evaluating the convolutions in the Wigner representation can be done by
systematically expanding the exponentials:\cite{rammersmith} 
\begin{equation}  \label{eq:43}
(A \circ B)(\varepsilon;t) = \mathrm{e}^{\mathrm{i} \left(
\partial^A_{\varepsilon} \partial^B_t - \partial^A_t
\partial^B_{\varepsilon} \right)/2} A(\varepsilon;t) B(\varepsilon;t),
\end{equation}
where the superscripts indicate which matrix the operator works on. A
significant simplification of the final result is achieved when the
superconducting gap is much less than the exchange energy, $\Delta \ll
\Delta_{xc}, E_F$. The energy dependence is then only determined by the
energy dependence of the Andreev reflection. Since we are evaluating the
energy gradients close to the Fermi level, $\partial_{\varepsilon} s_F \ll
\partial_{\varepsilon} r^A$, and we obtain the simplified expression for the
gradient matrix $A^{he}$: 
\begin{multline}  \label{eq:46}
A^{he}(\varepsilon;t) \approx - \frac{\mathrm{i}}{2} \partial_{\varepsilon}
\partial_t S^{he}_0 + \mathrm{i} t^*_{12} \partial_{\varepsilon} (M_e
r^A_{he}) \partial_t t_{21} \\
+ \mathrm{i} t^*_{12} \partial_{\varepsilon} \partial_t M_e r^A_{he} t_{21}
+ \mathrm{i} t^*_{12} \partial_t M_e \partial_{\varepsilon} M_e^{-1} M_e
r^A_{he} t_{21} \\
- \mathrm{i} t^*_{12} M_e r^A_{he} \partial_t r_{22} \partial_{\varepsilon}
r^A_{eh} r^*_{22} M_{e} r^A_{he} t_{21} \\
\equiv - \frac{\mathrm{i}}{2} \partial_{\varepsilon} \partial_t S_0^{he} +
\Gamma^{he}.
\end{multline}
Here, $S_0^{he}$ is the frozen scattering matrix from Eq.~(\ref{eq:16}), and
Before evaluating the currents, we observe that $\Gamma^{he}$ in the
rotating frame is diagonal in spin space. This fact, which is important when
evaluating non-equilibrium pumped charge and spin currents, can be seen from 
\begin{equation}  \label{eq:19}
\Gamma^{he} = \mathcal{U}\mathcal{V}^{\dagger} \underline{\Gamma}^{he} 
\mathcal{VU},
\end{equation}
with 
\begin{multline}  \label{eq:60}
\underline{\Gamma}^{he} = \frac{\mathrm{i}}{2} \underline{t}^*_{12}
\partial_{\varepsilon} (\underline{M}_e r^A_{he}) \Lambda ( \underline{t}%
_{21\uparrow} - \underline{t}_{21\downarrow}) \\
+ \frac{\mathrm{i}}{2} \underline{t}^*_{12} (\underline{M}_{e\uparrow} - 
\underline{M}_{e\downarrow}) \partial_{\varepsilon}(r^A_{he} \Lambda 
\underline{r}_{22} r^A_{eh}) \underline{r}^*_{22} \underline{M}_e r^A_{he} 
\underline{t}_{21} \\
- \frac{\mathrm{i}}{2} \underline{t}^*_{12} \underline{M}_e r^A_{he} (%
\underline{r}_{22\uparrow} - \underline{r}_{22\downarrow}) \Lambda
\partial_{\varepsilon} r^A_{eh} \underline{r}^*_{22} \underline{M}_e
r^A_{he} \underline{t}_{21} \\
- \frac{\mathrm{i}}{2} \underline{t}^*_{12} \partial_{\varepsilon} (%
\underline{M}_{e\uparrow} - \underline{M}_{e\downarrow}) r^A_{he} \Lambda 
\underline{t}_{21},
\end{multline}
where 
\begin{equation}  \label{eq:75}
\Lambda \equiv \mathcal{VU} \partial_t (\boldsymbol{m}\cdot\boldsymbol{\sigma%
}) \mathcal{U}^{\dagger} \mathcal{V}^{\dagger} = \partial_t \theta \sigma^x
+ \sin\theta \Omega \sigma^y.
\end{equation}
Multiplying $r^A_{he}$, which is $\sim \sigma^y$, with $\Lambda$, and using
that the other components in the equation are all diagonal, brings us to the
conclusion that $\underline{\Gamma}^{he}$ is diagonal in spin space.
Finally, we note that $\Gamma^{he} \to 0$ for a vanishing ferromagnetic
ordering parameter.

Once the gradient corrections to the frozen scattering matrix are derived,
one can obtain non-equilibrium pumped currents to first order in pumping
frequency.

\subsection{Pumped charge current}

\label{sec:pump-charge-curr}

According to Eq.~(\ref{eq:21}), the charge current is obtained by tracing
the matrix current~(\ref{eq:82}) over electron-hole, spin and mode space.
Making use of the electron-hole symmetries from Eqs.~(\ref{eq:4})(\ref{eq:10}%
), and using that both $\mathrm{Tr} \left\{ \partial_t S_0^{ee}
S_0^{ee\dagger} \right\} = 0$ and $\mathrm{Tr} \left\{ \partial_t S_0^{he}
S_0^{he\dagger} \right\} = 0$, one finds that the pumped charge current is
determined by 
\begin{multline}  \label{eq:1}
I_c (t) = \frac{e}{2\pi} \int_{-\infty}^{\infty} \mathrm{d}\varepsilon \: %
\biggr( \left[f_e (\varepsilon) - f_h (\varepsilon)\right] \mathrm{Tr} %
\Big\{ S_{0}^{he} S_{0}^{he\dagger} \\
+ A^{he} S_0^{he\dagger} + S_0^{he} A^{he\dagger} - \frac{1}{2} P \left\{
S_0^{he}; S_0^{he\dagger} \right\} \Big\} \biggr),
\end{multline}
to first order in pumping parameter frequency. Using that $A^{he} = - \frac{%
\mathrm{i}}{2} \partial_{\varepsilon} \partial_t S_0^{he} + \Gamma^{he}$,
the current~\eqref{eq:1} simplifies to 
\begin{multline}  \label{eq:2}
I_c (t) = \frac{e}{2\pi} \int_{-\infty}^{\infty} \mathrm{d}\varepsilon \: %
\biggr( \left[f_e (\varepsilon) - f_h (\varepsilon)\right] \\
\times \mathrm{Tr} \Big\{ S_{0}^{he} S_{0}^{he\dagger} + \Gamma^{he}
S_0^{he\dagger} + S_0^{he} \Gamma^{he\dagger} \Big\} \biggr).
\end{multline}
Any non-equilibrium pumped contributions to the current are determined by
the remainder $\Gamma^{he}$ from Eq.~(\ref{eq:60}). However, as pointed out
at the end of Sec.~\ref{sec:grad-corr-matr-1}, $\underline{\Gamma}^{he}$ is
a diagonal matrix in spin space. From Eq.~(\ref{eq:16}), we know that $%
\underline{S}_0^{he}$ is strictly off-diagonal in spin space. This implies
that $\mathrm{Tr} \left\{  \Gamma^{he} S_0^{he\dagger} \right\} = 0$, and
the charge current is reduced to the stationary result: 
\begin{equation}  \label{eq:3}
I_c = \frac{e}{2\pi} \int_{-\infty}^{\infty} \mathrm{d}\varepsilon \: \left[%
f_e (\varepsilon) - f_h (\varepsilon)\right] \tilde{g} (\varepsilon),
\end{equation}
where the total conductance is defined as 
\begin{equation}  \label{eq:47}
\tilde{g} \equiv \sum_{m,n} \left\{ \left|\underline{S}^{he}_{\downarrow%
\uparrow,mn} \right|^2 + \left|\underline{S}^{he}_{\uparrow\downarrow,mn}%
\right|^2 \right\}.
\end{equation}
The result in Eq.~(\ref{eq:3}) shows that there is no pumped charge current
in N|F|S structures, even when there is an additional bias voltage driving
the system, \textit{e.g.} there are no bilinear contributions proportional
to the bias voltage and the FMR frequency. The stationary result is similar
to that obtained in FS|N structures,\cite%
{brataas04:_spin_and_charg_pumpin_by} a result that indicates that the total
scattering matrix for a disordered region coupled to a ferromagnetic
superconductor, is structurally equivalent to that of a disordered
ferromagnetic region coupled to a superconductor. The two structures have
different scattering matrices, however, and therefore the expressions for
the conductances differ.

\subsection{Pumped spin current}

\label{sec:pumped-spin-current}

We proceed by evaluating the pumped spin current to first order in pumping
parameter frequency. Utilizing the electron-hole symmetry relations for the
total scattering matrix, we obtain 
\begin{widetext}
  \begin{multline}
    \label{eq:17}
    \boldsymbol{I}_s(t) = \frac{1}{4\pi} \int_{-\infty}^{\infty}
    \mathrm{d}\varepsilon\: \left(f_e(\varepsilon) -
      f_h(\varepsilon)\right) \left[\mathrm{Tr} \left\{
        \boldsymbol{\sigma}^* \left( S_0^{he} S_0^{he\dagger} +
          \Gamma^{he} S_0^{he\dagger} + S_0^{he} \Gamma^{he\dagger}
        \right) \right\} + \partial_{\varepsilon} \mathrm{Im}
      \mathrm{Tr} \left\{ \boldsymbol{\sigma}^* \partial_t S_0^{he}
        S_0^{he\dagger} \right\} \right] \\ + \frac{1}{4\pi}
    \int_{-\infty}^{\infty} \mathrm{d}\varepsilon \:
    (-\partial_{\varepsilon} f_e (\varepsilon)) \left[ \mathrm{Im}
      \mathrm{Tr} \left\{ \boldsymbol{\sigma} \partial_t S_0^{ee}
        S_0^{ee\dagger} \right\} - \mathrm{Im} \mathrm{Tr} \left\{
        \boldsymbol{\sigma}^* \partial_t S_0^{he}
        S_0^{he\dagger}\right\} \right].
  \end{multline}
\end{widetext}

Introducing the conductance polarization 
\begin{equation}  \label{eq:48}
\tilde{p} \equiv \frac{1}{\tilde{g}} \sum_{m,n} \left\{ \left|\underline{S}%
^{he}_{\downarrow\uparrow,mn} \right|^2 - \left|\underline{S}%
^{he}_{\uparrow\downarrow,mn}\right|^2 \right\},
\end{equation}
and the generalized mixing conductance\cite%
{brataas04:_spin_and_charg_pumpin_by} 
\begin{equation}  \label{eq:18}
\tilde{g}^{\uparrow\downarrow} \equiv \sum_{m,n} \left\{ \delta_{m,n} - 
\underline{S}^{ee}_{\uparrow,mn} \underline{S}^{ee*}_{\downarrow,mn} + 
\underline{S}^{he}_{\downarrow\uparrow,mn} \underline{S}^{he*}_{\uparrow%
\downarrow,mn} \right\}.
\end{equation}
we find the following expression for the spin current: 
\begin{widetext}
  \begin{multline}
    \label{eq:62}
    \boldsymbol{I}_s(t) = - \frac{1}{4\pi} \int_{-\infty}^{\infty}
    \mathrm{d}\varepsilon \: (f_e (\varepsilon) - f_h(\varepsilon))
    \left( \tilde{p}\tilde{g} \boldsymbol{m}(t) - \mathrm{Tr} \left\{
        \boldsymbol{\sigma}^*( \Gamma^{he} S_0^{he\dagger} + S_0^{he}
        \Gamma^{he\dagger}) \right\} \right) \\ + \frac{1}{8\pi}
    \int_{-\infty}^{\infty} \mathrm{d}\varepsilon \: (f_e
    (\varepsilon) - f_h(\varepsilon)) \partial_{\varepsilon} \biggr(
      \boldsymbol{m} \times \partial_t \boldsymbol{m}
      \Big(\tilde{g} + 2 \mathrm{Re} \sum_{m,n}
        \underline{S}^{he}_{\downarrow\uparrow,mn}
        \underline{S}^{he*}_{\uparrow\downarrow,mn} \Big) + 2
      \partial_t \boldsymbol{m} \mathrm{Im} \sum_{m,n}
      \underline{S}^{he}_{\downarrow\uparrow,mn}
      \underline{S}^{he*}_{\uparrow\downarrow,mn} \biggr) \\ +
      \frac{1}{4\pi} \int_{-\infty}^{\infty} \mathrm{d}\varepsilon \:
      \partial_{\varepsilon} f_e(\varepsilon) \left( \boldsymbol{m}
        \times \partial_t \boldsymbol{m} \mathrm{Re}
        \tilde{g}^{\uparrow\downarrow} + \partial_t \boldsymbol{m}
        \mathrm{Im} \tilde{g}^{\uparrow\downarrow} \right).
  \end{multline}
\end{widetext}

The term $\sim \tilde{p}\tilde{g}\boldsymbol{m}(t)$ on the right hand side
of Eq.~\eqref{eq:62} corresponds to the non-equilibrium bias voltage spin
current observed also in the absence of a precessing magnetization vector.
Terms in the final line are similar to those derived previously within
electro-chemical equilibrium pumping theory for F|N\cite{spinpumping}, and
FS|N structures\cite{brataas04:_spin_and_charg_pumpin_by}. However, we ask
the reader to note that the generalized mixing conductance in Eq.~(3) in
Ref.~\onlinecite{brataas04:_spin_and_charg_pumpin_by} is valid for triplet
superconductors only; the correct mixing conductance for a singlet
superconductor is given by Eq.~\eqref{eq:18} above. The remaining terms on
the right hand side of Eq.~\eqref{eq:62} are non-equilibrium, pumped
contributions to the spin current. They depend on pumping parameter
frequency via $\partial_t \boldsymbol{m}$ and the $\Lambda$ term from Eq.~(%
\ref{eq:75}), which is contained in the gradient remainder $\Gamma^{he}$.

Finally, we would like to point out that there are no pumped contributions
to the \emph{longitudinal} spin current $I^{||}_{s} \equiv \boldsymbol{m}
\cdot \boldsymbol{I}_s$. The terms in the second and third line of Eq.~(\ref%
{eq:62}) are transverse with respect to the magnetization $\boldsymbol{m}$,
so this leaves only a possible gradient remainder contribution coming from $%
\Gamma^{he}$. However, due to the particular matrix structure of $\Gamma^{he}
$ mentioned in Sec.~\ref{sec:grad-corr-matr-1}, $\boldsymbol{m} \cdot 
\mathrm{Tr}\{ \boldsymbol{\sigma}^* \Gamma^{he} S_0^{he\dagger} \}$
vanishes. This observation implies that the longitudinal spin current is
stationary and unaffected by the precessing magnetization. Thus, to first
order in precession frequency: 
\begin{equation}  \label{eq:41}
I_s^{||} = \boldsymbol{m}(t) \cdot \boldsymbol{I}_s (t) = - \frac{1}{4\pi}
\int_{-\infty}^{\infty} \mathrm{d} \varepsilon \: (f_e (\varepsilon) - f_h
(\varepsilon)) \tilde{p}\tilde{g}.
\end{equation}
In the following, we will investigate pumped charge and spin currents when
the ferromagnetic region is longer than the typical transverse spin
coherence length.

\subsection{Long ferromagnet limit}

\label{sec:limit-1:-long-2}

When the length $L_{F}$ of the ferromagnet is longer than the transverse
spin coherence length, 
\begin{equation}
L_{F}>\lambda _{F}\equiv \frac{\pi }{k_{F\uparrow }-k_{F\downarrow }},
\label{eq:39}
\end{equation}%
where $k_{F\sigma }$ is the Fermi wave vector of a spin $\sigma $ electron,
we expect to find a mixing conductance that is determined by the properties
of the N-F subsystem, characterized by the spin-dependent
conductances\cite{spinpumping} 
\begin{equation}
g^{\sigma \sigma ^{\prime }}=\sum_{m,n}\left( \delta _{m,n}-\underline{r}%
_{\sigma ,mn}\underline{r}_{\sigma ^{\prime },mn}^{\ast }\right) .
\label{eq:49}
\end{equation}%
Indeed, in the limit~(\ref{eq:39}), one can disregard \textquotedblleft
mixing transmission\textquotedblright\ terms, $\sum_{m,n}\underline{t}%
_{\sigma ,mn}\underline{t}_{-\sigma ,mn}^{\ast }\rightarrow 0$, so that $%
\sum_{m,n}\underline{S}_{\downarrow \uparrow ,mn}^{he}\underline{S}%
_{\uparrow \downarrow ,mn}^{he\ast }\rightarrow 0$. Disregarding
interference terms between reflected and transmitted electronic wave
functions, one obtains 
\begin{equation}
\sum_{m,n}\underline{S}_{\uparrow ,mn}^{ee}\underline{S}_{\downarrow
,mn}^{ee\ast }\rightarrow \sum_{m,n}\underline{r}_{11\uparrow ,mn}\underline{%
r}_{11\downarrow ,mn}^{\ast },  \label{eq:45}
\end{equation}%
for a long ferromagnet. This implies that $\tilde{g}^{\uparrow \downarrow
}\rightarrow g^{\uparrow \downarrow }$, while the total conductance $\tilde{g%
}$ and the conductance polarization $\tilde{p}$ remain unchanged. Since the
mixing conductance is now determined by properties of the N-F structure, energy gradients of the mixing conductance should be
disregarded in the limit $\Delta \ll \Delta _{xc},E_{F}$, as described in
Sec.~\ref{sec:grad-corr-matr-1}. Finally, by an explicit calculation, one
can show that $\mathrm{Tr}\{\Gamma ^{he}S_{0}^{he\dagger }\boldsymbol{\sigma 
}^{\ast }\}\sim \underline{t}_{\sigma }\underline{t}_{-\sigma }^{\ast }$,
which vanishes when Eq.~(\ref{eq:39}) holds. To summarize, when the
ferromagnet is longer than the transverse spin coherence length, the charge
current and longitudinal spin current are still given by 
\begin{equation}
I_{c}=\frac{e}{2\pi }\int_{-\infty }^{\infty }\mathrm{d}\varepsilon \left(
f_{e}(\varepsilon )-f_{h}(\varepsilon )\right) \tilde{g},  \label{eq:11}
\end{equation}%
and 
\begin{equation}
I_{s}^{||}=-\frac{1}{4\pi }\int_{-\infty }^{\infty }\mathrm{d}\varepsilon
(f_{e}(\varepsilon )-f_{h}(\varepsilon ))\tilde{p}\tilde{g},  \label{eq:55}
\end{equation}%
while the transverse spin current is simplified to 
\begin{multline}
\boldsymbol{I}_{s}^{\perp }(t)=-\frac{1}{8\pi }\int_{-\infty }^{\infty }%
\mathrm{d}\varepsilon (\partial _{\varepsilon }f_{e}(\varepsilon )-\partial
_{\varepsilon }f_{h}(\varepsilon ))\tilde{g}\boldsymbol{m}\times \partial
_{t}\boldsymbol{m}  \label{eq:12} \\
+\frac{1}{4\pi }\int_{-\infty }^{\infty }\mathrm{d}\varepsilon \partial
_{\varepsilon }f_{e}(\varepsilon )\left( \mathrm{Re}{g}^{\uparrow \downarrow
}\boldsymbol{m}\times \partial _{t}\boldsymbol{m}+\mathrm{Im}{g}^{\uparrow
\downarrow }\partial _{t}\boldsymbol{m}\right) .
\end{multline}%
With no applied bias voltage, the pumped spin current in Eq.~(\ref{eq:12})
is identical to that found in N-F systems\cite{spinpumping}%
, as should be expected. In this situation, emission of spins from the
ferromagnet into the normal metal are unaffected by the superconductor. 
\begin{figure}[th]
\centering 
\subfigure[Transverse spin component $\partial_t
  \boldsymbol{m} \cdot \boldsymbol{I}_s/\Omega$]{\scalebox{0.74}{
    \begin{picture}(0,0)      \includegraphics{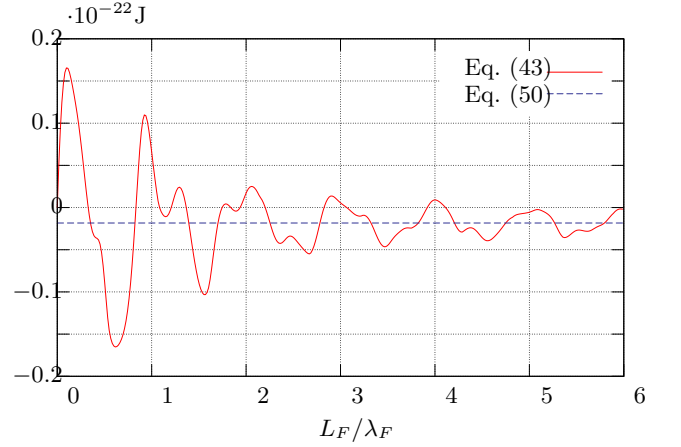}
    \end{picture}}  \begingroup
  \setlength{\unitlength}{0.0200bp}  \begin{picture}(11840,8208)(0,0)    \put(722,1221){\makebox(0,0)[r]{\strut{}$-0.2$}}    \put(722,2812){\makebox(0,0)[r]{\strut{}$-0.1$}}    \put(722,4403){\makebox(0,0)[r]{\strut{}$0$}}    \put(722,5994){\makebox(0,0)[r]{\strut{}$0.1$}}    \put(722,7585){\makebox(0,0)[r]{\strut{}$0.2$}}    \put(800,8000){\makebox(0,0)[l]{\strut{}$\cdot 10^{-22}
        \text{J}$}}    \put(925,814){\makebox(0,0){\strut{}$0$}}    \put(2700,814){\makebox(0,0){\strut{}$1$}}    \put(4500,814){\makebox(0,0){\strut{}$2$}}    \put(6280,814){\makebox(0,0){\strut{}$3$}}    \put(8050,814){\makebox(0,0){\strut{}$4$}}    \put(9850,814){\makebox(0,0){\strut{}$5$}}    \put(11600,814){\makebox(0,0){\strut{}$6$}}    \put(6262,204){\makebox(0,0){\strut{}$L_F/\lambda_F$}}    \put(9953,7006){\makebox(0,0)[r]{\strut{}Eq.~(\ref{eq:62})}}    \put(9953,6499){\makebox(0,0)[r]{\strut{}Eq.~(\ref{eq:12})}}  \end{picture}  \endgroup}
\\[0.25cm]
\subfigure[Transverse spin component $(\boldsymbol{m} \times
\partial_t \boldsymbol{m}) \cdot
\boldsymbol{I}_s/\Omega$]{\scalebox{0.74}{
    \begin{picture}(0,0)      \includegraphics{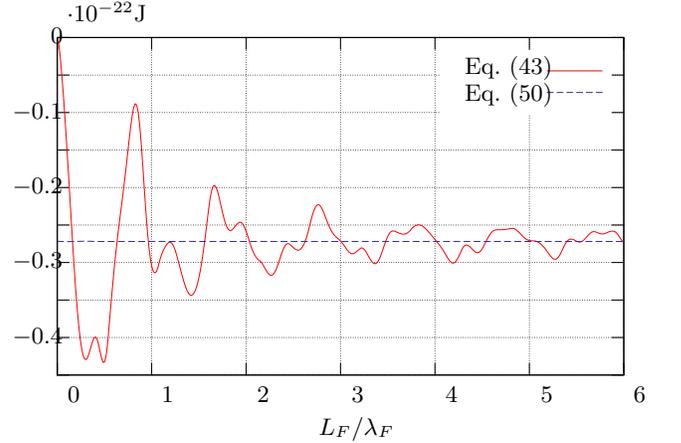}
    \end{picture}}  \begingroup
  \setlength{\unitlength}{0.0200bp}  \begin{picture}(11840,8208)(0,0)    \put(722,1928){\makebox(0,0)[r]{\strut{}$-0.4$}}    \put(722,3343){\makebox(0,0)[r]{\strut{}$-0.3$}}    \put(722,4757){\makebox(0,0)[r]{\strut{}$-0.2$}}    \put(722,6171){\makebox(0,0)[r]{\strut{}$-0.1$}}    \put(722,7585){\makebox(0,0)[r]{\strut{}$0$}}    \put(800,8000){\makebox(0,0)[l]{\strut{}$\cdot 10^{-22}
        \text{J}$}}    \put(925,814){\makebox(0,0){\strut{}$0$}}    \put(2700,814){\makebox(0,0){\strut{}$1$}}    \put(4500,814){\makebox(0,0){\strut{}$2$}}    \put(6280,814){\makebox(0,0){\strut{}$3$}}    \put(8050,814){\makebox(0,0){\strut{}$4$}}    \put(9850,814){\makebox(0,0){\strut{}$5$}}    \put(11600,814){\makebox(0,0){\strut{}$6$}}    \put(6262,204){\makebox(0,0){\strut{}$L_F/\lambda_F$}}    \put(9953,7006){\makebox(0,0)[r]{\strut{}Eq.~(\ref{eq:62})}}    \put(9953,6499){\makebox(0,0)[r]{\strut{}Eq.~(\ref{eq:12})}}  \end{picture}  \endgroup}
\caption{Exact (red line) and approximate (blue dashed line) transverse spin
currents for a ballistic N-F-S structure
as functions of length of the ferromagnetic region. In the plot, $E_{F}=10$
eV, $\Delta _{xc}=9E_{F}/16$, $\Delta =E_{F}/160$, $eV=\Delta /2$ and $%
\Omega =0.2$ GHz.}
\label{fig:plot1}
\end{figure}

To compare the exact result~(\ref{eq:62}) with the long ferromagnet
approximation of Eq.~(\ref{eq:12}), we plot in Fig.~\ref{fig:plot1} the spin
current along $\partial_t\boldsymbol{m}$ for a ballistic N|F|S trilayer, as
a function of the ratio between the ferromagnet length ($L_F$) and the
transverse spin coherence length ($\lambda_F$) defined in Eq.~(\ref{eq:39}).
When $L_F \le \lambda_F$, non-negligible ``mixing transmission'' terms
combine with energy gradients of the scattering matrix and produce large
deviations between the two equations. As $L_F$ exceeds $\lambda_F$, the fit
improves and the exact result oscillates towards the spin current obtained
by the approximate Eq.~\eqref{eq:12}.

\section{Conclusion}

\label{sec:conclusion}

In conclusion, we have derived non-equilibrium pumped charge and spin
currents to first order in pump frequency, using time-dependent scattering
theory. Magnetization precession induces transverse spin currents, but
neither charge nor longitudinal spin currents, which are both given by their
stationary values. The currents are expressed in terms of generalized, spin
dependent conductances, that include spin-dependent scattering in the
ferromagnet and Andreev reflection at the F|S interface. Finally, we
consider trilayers where the ferromagnetic region is longer than the
transverse spin coherence length, and derive an approximate expression for
the transverse spin current. Numerical calculation of the spin current in a
ballistic trilayer shows good agreement between exact and approximate spin
currents for ferromagnets whose layer thicknesses exceed the transverse spin
coherence length.



\end{document}